\documentclass[12pt,a4paper]{article}

\usepackage[utf8]{inputenc}
\usepackage[T1]{fontenc}
\usepackage{mathpazo}        
\usepackage{amsmath,amssymb,amsthm}
\usepackage{graphicx}
\usepackage{booktabs}
\usepackage{hyperref}
\usepackage{url}
\usepackage[margin=1in]{geometry}
\usepackage{tikz}
\usepackage{pgfplots}
\pgfplotsset{compat=1.18}
\usepackage{xcolor}
\usepackage{enumitem}
\usepackage{caption}
\usepackage{subcaption}
\usepackage{float}
\usepackage{cite}
\usepackage{fancyhdr}

\definecolor{coral}{RGB}{220,80,60}
\definecolor{deepblue}{RGB}{25,55,109}
\definecolor{olive}{RGB}{85,107,47}
\definecolor{gold}{RGB}{200,140,50}

\newtheorem{definition}{Definition}[section]

\newtheorem{proposition}{Proposition}[section]

\title{\textbf{Agentic Software:}\\
       \large How AI Agents Are Restructuring the Software Paradigm}

\author{%
  Zhenfeng Cao\\
  \small Lingxi Intelligent Investment (Shenzhen) Development Co., Ltd.\\
  \small \texttt{info@stellarsea.com}
}

\date{\today}

\begin{document}

\maketitle

\begin{abstract}
For over half a century, software engineering has operated on a foundational premise: human engineers decompose problems, encode decision logic into static code, and manually adapt that code as requirements evolve. This paper argues that the emergence of AI agents -- systems where large language models serve as the primary reasoning engine, dynamically generating and discarding code as an instrumental resource -- constitutes a fundamental restructuring of what software is, not an incremental tool improvement. We formalize the distinction between traditional deterministic software and agentic software: in the former, code is the carrier of pre-written decision logic; in the latter, the agent itself is the software, and its decision logic is generated at runtime. We trace the historical arc from licensed software to SaaS to Agent-as-a-Service (AaaS), showing that each shift transferred additional complexity away from end-users -- with the agentic shift transferring not just operational complexity but decision-making complexity itself. We introduce Agentic Engineering as an expansion of the software engineering discipline into a new paradigm, distinct in its core object of study (agent systems rather than static source code), its control model (LLM-driven rather than human-predefined), and its human role (intent architect rather than code author). Through analysis of recent benchmark evidence including SWE-bench Verified, EvoClaw, and LangChain's multi-agent coordination studies, we demonstrate both the transformative potential of the agentic paradigm and its current limitations. We conclude with a four-stage roadmap toward self-evolving agent ecosystems and concrete recommendations for practitioners navigating this transition.
\end{abstract}

\section{Introduction}

Software engineering, as codified at the 1968 NATO Conference \cite{nato1968}, was born from a crisis: systems were growing in complexity beyond what ad-hoc programming practices could manage. The discipline's founding insight was that rigorous methodologies---structured design, modular decomposition, configuration management, systematic testing---could tame this complexity. For five decades, this bet largely paid off. We moved from waterfall to agile, from monoliths to microservices, from manual deployment to CI/CD.

Yet a deeper structural problem persisted. As Brooks observed in \textit{The Mythical Man-Month} \cite{brooks1975}, software complexity exhibits a fundamentally different scaling behavior than other engineering domains. Unlike bridges or circuits, software has no manufacturing step---the design \textit{is} the product. Every new feature, every edge case, every integration point adds to a combinatorial explosion of possible states and interactions that Brooks characterized as ``essential complexity'': complexity inherent to the problem itself, not accidental to the implementation.

This paper contends that the emergence of AI agents does not merely offer a new tool within the existing paradigm. Rather, it fundamentally restructures the premise on which software engineering was founded -- not by ending it, but by expanding its definition. An AI agent is itself software: it runs on hardware, processes data, and produces outputs. What distinguishes it is not that it escapes the category of software, but that it represents a new kind of software -- one where decision logic is no longer pre-written but dynamically generated at runtime. When a large language model (LLM) \cite{brown2020language} can understand a task, decompose it into subtasks, dynamically generate code to execute those subtasks, and discard that code when it's no longer needed, the role of code changes from \textit{the system itself} to \textit{an ephemeral instrument of reasoning}. This shift is as fundamental as the transition from analog circuits to stored-program computers.

We make three central claims:
\begin{enumerate}[leftmargin=*]
    \item \textbf{First-Principles Necessity.} The agentic paradigm is not a market preference but an inevitable consequence of complexity scaling laws. Traditional software requires human engineers to explicitly encode every decision; LLM-based agents can navigate complexity non-linearly by outsourcing reasoning to models whose capacity grows with training compute.
    \item \textbf{Software Redefined, Not Replaced.} The transition from ``AI $\rightarrow$ Software $\rightarrow$ Result'' to ``Agent $\rightarrow$ Result'' does not eliminate software -- the agent itself is software, albeit of a fundamentally different kind. Rather, it collapses the intermediary: the agent is simultaneously the software system and its execution engine, obviating the need for a separate, statically coded artifact. We formalize this as the third major paradigm shift in what software is and how it is delivered.
    \item \textbf{Emergent Discipline.} Agentic Engineering is emerging as a distinct practice with its own concepts, tools, and metrics. Its practitioners are not ``better programmers'' but a fundamentally different role: intent architects, agent coordinators, and outcome auditors.
\end{enumerate}

The remainder of this paper is structured as follows. Section 2 presents a first-principles analysis of traditional software and agent-based systems, including a formal complexity argument. Section 3 traces the historical paradigm shifts in software delivery and positions AaaS as the logical endpoint. Section 4 defines Agentic Engineering as a discipline and contrasts it with traditional software engineering. Section 5 reviews empirical evidence from recent benchmarks, acknowledging both breakthroughs and persistent challenges. Section 6 proposes an evolutionary roadmap. Section 7 concludes with implications for practitioners and the research community.

\section{First-Principles Analysis}

\subsection{The Nature of Traditional Software}

We begin with a precise definition.

\begin{definition}[Traditional Software System]
A traditional software system $S$ is a tuple $S = (C, D, E)$ where:
\begin{itemize}
    \item $C$ is a set of computational resources (CPU, memory, I/O);
    \item $D$ is a set of deterministic decision rules encoded in source code;
    \item $E$ is an execution environment that evaluates $D$ against inputs to produce outputs.
\end{itemize}
The critical property is that $D$ is \textit{static with respect to execution}: all decision logic must be explicitly written by human engineers before the system encounters any input.
\end{definition}

Under this definition, every feature addition, every bug fix, every adaptation to a changing environment requires a human to (a) understand the change needed, (b) locate the correct position in $D$, (c) modify the logic without introducing regressions, and (d) verify correctness. The cost of each change is a function of the size of $D$ and the density of its internal dependencies.

\subsection{The Complexity Barrier}

Brooks \cite{brooks1975} distinguished between \textit{accidental complexity} (artifacts of particular implementations) and \textit{essential complexity} (inherent to the problem). While decades of advances---higher-level languages, frameworks, automated testing---have systematically reduced accidental complexity, essential complexity remains unbounded. In fact, as systems grow, the interaction surface between components grows combinatorially.

\begin{proposition}[Complexity Scaling]
For a system with $n$ components, each potentially interacting with any other, the number of possible interaction topologies is $2^{\binom{n}{2}}$, which grows as $\Theta(2^{n^2})$. While real systems do not realize all configurations, the upper bound on complexity grows super-exponentially, while human cognitive capacity to reason about these interactions is essentially constant.
\end{proposition}

This mismatch is the deep structural reason why software projects experience declining marginal productivity as they grow. The traditional response---hierarchical decomposition, modular interfaces, encapsulation---reduces the constant factor but does not change the asymptotic behavior.

\subsection{Agentic Systems: A Formal Model}

In contrast, an agentic system operates on fundamentally different principles.

\begin{definition}[AI Agent System]
An AI agent system $A$ is a tuple $A = (M, \mathcal{T}, \mathcal{M}, \Pi)$ where:
\begin{itemize}
    \item $M$ is a large language model serving as the \textit{reasoning engine};
    \item $\mathcal{T}$ is a set of executable tools (code interpreters, APIs, databases, file systems);
    \item $\mathcal{M}$ is a memory subsystem (short-term context, long-term vector store);
    \item $\Pi$ is a planning mechanism that decomposes user intent into action sequences.
\end{itemize}
The system operates by iteratively executing: $a_t \leftarrow M(s_t, \mathcal{M})$, $s_{t+1} \leftarrow \text{exec}(a_t)$, where $s_t$ is the system state at time $t$ and $a_t$ is the action chosen by the model.
\end{definition}

The key distinction is that in an agentic system, the decision logic is \textit{generated at runtime}. The LLM $M$ can dynamically produce code, invoke tools, and adjust its behavior based on intermediate results---none of which was explicitly pre-programmed. The code it generates is not the system; it is a transient artifact, produced and discarded as needed.

This distinction maps cleanly to Karpathy's ``Software 2.0'' framework \cite{karpathy2017}, but extends it further. In Karpathy's formulation, neural networks replace hand-crafted program logic with learned weights. Agentic systems go a step further: the neural network does not merely \textit{replace} the program---it \textit{writes programs on demand}, using code as a tool in service of broader reasoning goals. This pattern is consistent with the ReAct framework \cite{yao2023react}, which demonstrated that interleaving reasoning traces with tool-use actions substantially improves task performance, and with Chain-of-Thought prompting \cite{wei2022chain}, which showed that explicit intermediate reasoning steps unlock latent capabilities in LLMs.

\subsection{Why the Agentic Paradigm Scales Differently}

Consider a task $T$ whose solution requires reasoning over a space of size $N$. Under the traditional paradigm:
\begin{itemize}
    \item A human engineer must mentally traverse this space to identify the solution path.
    \item The path must then be encoded as a static program.
    \item Human cognitive capacity $C_H$ is essentially fixed.
    \item Thus, for $N > C_H$, the task is infeasible at any realistic cost.
\end{itemize}

Under the agentic paradigm:
\begin{itemize}
    \item The LLM $M$ traverses the space, with effective capacity $C_M$ that scales with model size and training compute.
    \item The plan $\Pi$ decomposes $T$ into subproblems, each handled independently.
    \item Code is generated only for the specific solution path, not for all contingencies.
    \item As LLM capabilities improve (which they have been, exponentially), $C_M$ grows correspondingly.
\end{itemize}

Thus, the agentic paradigm decouples solution capacity from human cognitive limits. This is not a 10\% improvement; it is a qualitative change in what kinds of problems can be economically addressed.

\section{From SaaS to AaaS: The Third Paradigm Shift\texorpdfstring{}{}}

\subsection{Three Generations of Software Delivery}

The history of commercial software can be understood as a progressive transfer of complexity away from the end-user. Table~\ref{tab:paradigms} summarizes this trajectory.

\begin{table}[H]
\centering
\caption{Three Generations of Software Delivery}
\label{tab:paradigms}
\begin{tabular}{@{}p{2cm} p{2.8cm} p{2.5cm} p{2.8cm} p{2.5cm}@{}}
\toprule
\textbf{Generation} & \textbf{Core Mechanism} & \textbf{Complexity Owner} & \textbf{Revenue Model} & \textbf{Exemplars} \\
\midrule
Software 1.0 (Local) & Code + data execute on-premise & End-user (installation, maintenance) & License sale & Microsoft, Oracle \\
Software 2.0 (SaaS) & Code + data execute in cloud & Vendor (infrastructure, updates) & Subscription & Salesforce, AWS \\
Software 3.0 (AaaS) & Agent autonomously operates in cloud & Agent (understanding, building, running) & Outcome-based & OpenAI, Anthropic \\
\bottomrule
\end{tabular}
\end{table}

Each transition follows the same pattern: the party best positioned to absorb complexity absorbs it, and the party least positioned to manage it is liberated from it. SaaS liberated businesses from server rooms; AaaS promises to liberate them from the need to specify \textit{how} a result should be produced---they need only specify \textit{what} result they want.

\subsection{The Failure of ``AI $\rightarrow$ Software $\rightarrow$ Result''}

The dominant enterprise AI paradigm to date has been \textit{AI-augmented development}: use LLMs to help human engineers write code faster, within the traditional software lifecycle. We denote this as the ``AI $\rightarrow$ Software $\rightarrow$ Result'' pipeline.

This approach has three structural weaknesses:

\begin{enumerate}[leftmargin=*]
    \item \textbf{Bottleneck persistence.} The human engineer remains the critical path for design decisions, architecture, integration testing, and deployment. AI accelerates code generation (a sub-step of implementation) but does not remove the human from any phase.
    
    \item \textbf{Complexity ceiling intact.} The final deliverable remains a traditional software system $S = (C, D, E)$. Its complexity still scales with the size of $D$, and it still requires human understanding for any modification. AI merely made construction of $D$ somewhat faster.
    
    \item \textbf{Iteration latency.} Even with AI assistance, any functional change requires traversing the full chain: requirements $\rightarrow$ design $\rightarrow$ code $\rightarrow$ test $\rightarrow$ deploy. This latency cannot be reduced below human communication and coordination speeds.
\end{enumerate}

\subsection{``Agent $\rightarrow$ Result'': The Agent as Software}

The alternative paradigm collapses the distinction between software and its execution: the agent is the software. Rather than a human-built static system that produces results, an agent is a dynamic system that reasons, generates code, executes it, and delivers outcomes -- all within a single integrated loop.

\begin{enumerate}[leftmargin=*]
    \item Human articulates intent and constraints to an agent.
    \item Agent autonomously plans, executes (generating code as needed), validates, and delivers the result.
    \item Human audits the outcome and provides feedback.
\end{enumerate}

In this model, the agent is both the software and its operator. It may generate thousands of lines of code, execute database queries, call external APIs, produce visualizations -- all ephemerally, in service of the outcome. What persists is not the intermediate code but the agent's capability. As Kumar and Ramagopal \cite{kumar2026agentic} put it: ``AI coding agents excel at translating intent into code within a single user-driven session. Agentic engineering operates at a higher level of abstraction -- it's a control plane that orchestrates cross-team workflows, maintains long-term memory across agents, and manages state and traceability across the full software delivery lifecycle.''

\section{Agentic Engineering: Expanding the Discipline}

\subsection{Defining the Field}

Agentic Engineering, formally introduced by LangChain in April 2026 \cite{kumar2026agentic}, is defined as ``a multi-agent coordination model where AI agents function as digital team members---each with defined roles, shared memory, and a unified observability layer---to drive software through the entire delivery pipeline, not merely to generate code faster.'' We argue that Agentic Engineering does not replace software engineering but expands it: the agent itself is software, and building, deploying, and governing agent systems is simply the next frontier of the discipline.

Wang et al. \cite{wang2024agents} provide a foundational taxonomy of LLM-based agents in software engineering, identifying three core modules; a complementary survey by Guo et al.\ \cite{guo2024multiagent} offers a systematic treatment of multi-agent collaboration patterns and progress in LLM-based multi-agent systems.

\begin{figure}[H]
\centering
\begin{tikzpicture}[
    box/.style={draw, rounded corners, minimum width=3cm, minimum height=1.2cm, align=center, fill=white},
    arrow/.style={->, >=stealth, thick}
]
    \node[box, fill=blue!5, draw=deepblue] (perception) at (0,0) {\textbf{Perception}\\\small Multi-modal input\\\small processing};
    \node[box, fill=blue!5, draw=deepblue] (memory) at (4,0) {\textbf{Memory}\\\small Semantic, episodic,\\\small procedural};
    \node[box, fill=blue!5, draw=deepblue] (action) at (8,0) {\textbf{Action}\\\small Internal reasoning +\\\small external tool use};
    
    \node[box, fill=coral!8, draw=coral, minimum width=8cm, minimum height=1cm] (llm) at (4,2.5) {\textbf{LLM Reasoning Core}};
    
    \node[box, fill=olive!5, draw=olive, minimum width=6cm] (env) at (4,-2.5) {\textbf{External Environment}};
    
    \draw[arrow, deepblue] (llm.south) -- (perception.north);
    \draw[arrow, deepblue] (llm.south) -- (memory.north);
    \draw[arrow, deepblue] (llm.south) -- (action.north);
    \draw[arrow] (perception.south) -- (env.north west);
    \draw[arrow] (action.south) -- (env.north east);
    \draw[arrow] (env.north) -- (4,-1.2) -- (memory.south);
    
\end{tikzpicture}
\caption{The LLM-based agent framework for software engineering, adapted from Wang et al. \cite{wang2024agents}. The perception module handles multi-modal input; the memory module maintains semantic, episodic, and procedural knowledge; the action module executes both internal reasoning and external tool invocations. All are orchestrated by the LLM reasoning core.}
\label{fig:agent_framework}
\end{figure}
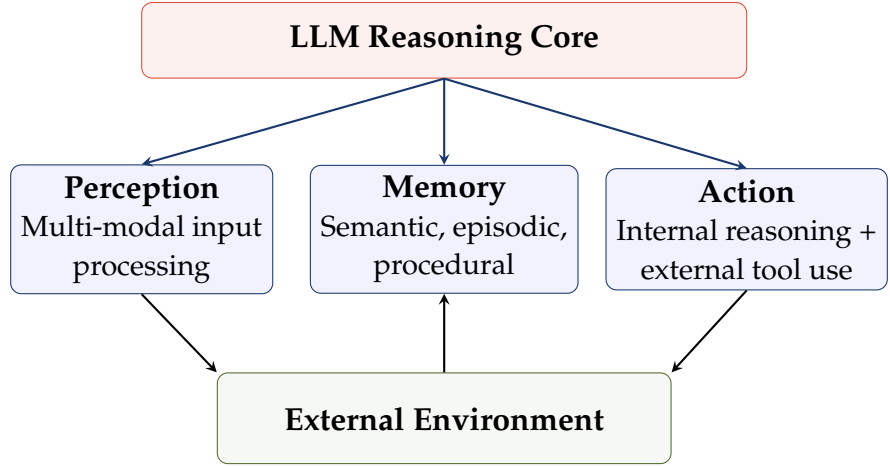

A concrete realization of this architecture can be observed in Hermes Agent \cite{hermes2026}, an open-source framework by Nous Research that operationalizes the perception-memory-action model with a distinctive self-evolution mechanism. Its most consequential feature is a closed learning loop: after completing complex tasks, the agent autonomously creates reusable Skills---parameterized procedural modules---that self-improve during subsequent use, automatically patching themselves when found insufficient. Cross-session episodic memory is realized through FTS5-backed conversation search with LLM summarization, enabling the agent to accumulate experiential knowledge over time. The framework's subagent delegation mechanism further demonstrates early multi-agent coordination in a widely deployed production system.

\subsection{Contrasting Agentic and Traditional Engineering}

Table~\ref{tab:contrast} maps the key dimensions of difference between the two paradigms.

\begin{table}[H]
\centering
\caption{Traditional Software Engineering vs.\ Agentic Engineering}
\label{tab:contrast}
\begin{tabular}{@{}p{3.2cm} p{5.2cm} p{5.2cm}@{}}
\toprule
\textbf{Dimension} & \textbf{Traditional SE} & \textbf{Agentic Engineering} \\
\midrule
Core artifact & Source code (static) & Agent system (dynamic) \\
Control center & Human engineer & LLM reasoning engine \\
Decision mechanism & Pre-designed logic & Runtime-generated reasoning \\
Development cycle & Linear (design$\rightarrow$code$\rightarrow$test) & Autonomous iterative loop \\
Human role & Code author & Intent architect, coordinator, auditor \\
Complexity ceiling & Human cognition ($O(1)$) & Model capacity (growing with compute) \\
\midrule
Output unit & Functioning software & Delivered outcomes \\
Error handling & Programmer-defined & Model-adaptive \\
Evolution & Manual refactoring & Self-modification \\
\bottomrule
\end{tabular}
\end{table}

\subsection{The Human Role Reimagined}

Perhaps the most consequential shift is in the human role. In the traditional paradigm, human value was measured by the ability to produce correct, efficient code. In the agentic paradigm, code-generation skill becomes commoditized. The new human differentiators are:

\begin{itemize}[leftmargin=*]
    \item \textbf{Intent articulation.} The ability to specify goals with sufficient clarity and constraint that agents can operate autonomously without producing unintended outcomes.
    \item \textbf{Architectural oversight.} Understanding at the system level how multiple agents should coordinate, what memory should be shared, and where human judgment must intervene.
    \item \textbf{Quality calibration.} Defining what ``good'' looks like and building evaluation frameworks that agents can use for self-correction.
    \item \textbf{Ethical governance.} Ensuring agent behavior aligns with organizational values, legal requirements, and societal expectations.
\end{itemize}

We believe the implications for individual practitioners are profound: as agentic capabilities mature, the productivity multiplier for those who master agent orchestration will far exceed the traditional ``10x engineer'' benchmark---not through faster typing, but through the ability to coordinate swarms of agents toward complex outcomes. The ceiling is not fixed; it rises with each advance in model capability and orchestration infrastructure.

\section{Empirical Evidence and Current Limitations}

\subsection{Breakthrough Results}

The empirical record provides strong evidence for the agentic thesis. We highlight four representative data points.

\textbf{SWE-bench Verified.} Ma et al.\ \cite{ma2024swegpt} demonstrated that Lingma SWE-GPT 72B, an open development-process-centric model, resolves 30.20\% of GitHub issues on SWE-bench Verified---approaching GPT-4o's 31.80\% while being fully open. Notably, even the 7B variant resolved 18.20\%, proving that small models can perform meaningful automated software engineering when trained on process data rather than static code alone. This represents a 22.76\% relative improvement over Llama 3.1 405B, a model nearly 6$\times$ larger.

\textbf{Multi-Agent Coordination.} Kumar and Ramagopal \cite{kumar2026agentic} report results from a pilot study deploying coordinated agent swarms across 20+ enterprise debugging workflows. The coordinated agent system reduced root-cause identification time by 93\%, saving over 200 engineering hours in a single month. Critically, these gains came not from better individual agents but from \textit{orchestration}---the ability to maintain shared context across agents, to parallelize investigation, and to cross-validate findings.

\textbf{Self-Evolution.} Hermes Agent \cite{hermes2026}, an open-source framework by Nous Research with over 179,000 GitHub stars, provides the most complete realization of the self-evolution principle in a production system. Its architecture implements a closed learning loop: after completing complex tasks, the agent autonomously creates reusable ``Skills''---parameterized procedural modules that capture successful strategies. Critically, these skills self-improve during use---when a skill is invoked and found lacking, the agent patches it automatically, accumulating refinements over successive interactions. This pattern---create, use, detect weakness, self-patch---operates without human intervention, embodying precisely the self-evolution dynamic that distinguishes agentic systems from traditional software. Cross-session continuity is maintained through FTS5-backed conversation search with LLM summarization, enabling the agent to recall and build upon prior experiences. The framework's subagent delegation mechanism further demonstrates early multi-agent coordination in a widely deployed system.

\textbf{Generalization.} Wang et al.\ \cite{wang2024agents} catalog hundreds of studies applying LLM-based agents across the full software lifecycle: requirements analysis, architecture design, code generation, testing, debugging, deployment, and maintenance. The breadth of coverage suggests that the agentic pattern is not limited to narrow tasks but generalizes across software engineering activities.

\subsection{Persistent Challenges}

Despite rapid progress, significant challenges remain. The EvoClaw benchmark \cite{deng2026evoclaw} provides the most sobering data. Deng et al.\ constructed a benchmark requiring agents to perform \textit{continuous} software evolution---not isolated issue fixes but sustained development across commit histories, where each change must preserve system integrity and where errors accumulate. Their key finding:

\begin{quote}
\textit{``Overall performance scores drop significantly from $>80\%$ on isolated tasks to at most 38\% in continuous settings, exposing agents' profound struggle with long-term maintenance and error propagation.''} \cite{deng2026evoclaw}
\end{quote}

This reveals four core challenges:

\begin{enumerate}[leftmargin=*]
    \item \textbf{Context drift.} As codebases grow beyond the effective context window, agents lose coherent understanding of system-wide invariants and dependencies.
    \item \textbf{Error propagation.} A small error in an early commit cascades into compounding failures in subsequent work, and agents lack robust mechanisms for detecting and recovering from these chains.
    \item \textbf{Technical debt awareness.} Agents do not currently model the long-term costs of their design decisions---they optimize for immediate task completion without considering maintainability.
    \item \textbf{Verification fidelity.} Automated testing remains incomplete; agents can pass tests while introducing subtle semantic errors that only manifest under novel inputs.
\end{enumerate}

\begin{figure}[H]
\centering
\begin{tikzpicture}
\begin{axis}[
    width=0.85\textwidth,
    height=6cm,
    ybar,
    bar width=25pt,
    ymin=0, ymax=100,
    ylabel={Success Rate (\%)},
    xlabel={Evaluation Setting},
    symbolic x coords={Isolated Tasks,Continuous Evolution},
    xtick=data,
    nodes near coords,
    nodes near coords align={vertical},
    enlarge x limits=0.3,
    title={Performance Degradation in Continuous Evolution (EvoClaw)},
    title style={font=\small}
]
\addplot[fill=deepblue!30, draw=deepblue] coordinates {(Isolated Tasks,82) (Continuous Evolution,38)};
\draw[->, thick, coral] (axis cs:Isolated Tasks,55) -- (axis cs:Continuous Evolution,55) node[midway, above] {\small \textbf{--54\% drop}};
\end{axis}
\end{tikzpicture}
\caption{Agent performance on the EvoClaw benchmark \cite{deng2026evoclaw}. When evaluated on continuous software evolution (requiring sustained development across commits with error accumulation), success rates collapse from over 80\% to at most 38\%. Data based on evaluation of 12 frontier models across 4 agent frameworks.}
\label{fig:evoclaw}
\end{figure}
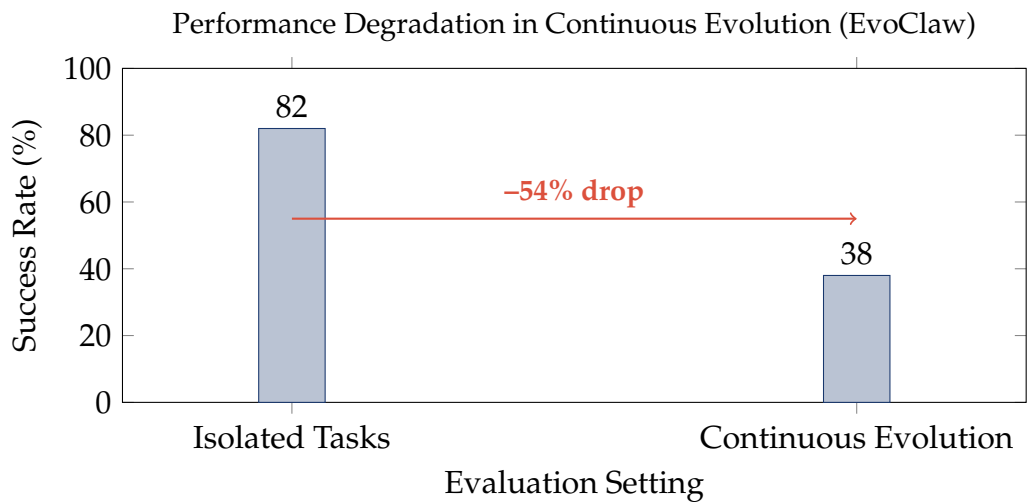

Figure~\ref{fig:evoclaw} visualizes the performance cliff that EvoClaw reveals.

\subsection{The Gap Analysis}

The gap between isolated-task performance ($>80\%$) and continuous-evolution performance ($<38\%$) quantifies the distance between current agent capability and the threshold for fully autonomous software engineering. This gap is not fundamental---it reflects limitations in context management, memory architecture, and verification mechanisms that are active areas of research. But it serves as an important calibration: agentic engineering is real and transformative today as an \textit{augmentation} paradigm, but will require several more years of concentrated research before fully autonomous software development becomes reliable in production settings.

\section{Evolutionary Roadmap}

Based on current capabilities and trajectories, we propose a four-stage roadmap for the evolution of agentic engineering. Table~\ref{tab:roadmap} summarizes.

\begin{table}[H]
\centering
\caption{Four-Stage Evolution of Agentic Engineering}
\label{tab:roadmap}
\small
\begin{tabular}{@{}p{1.5cm} p{2.5cm} p{3cm} p{2.5cm} p{2.5cm}@{}}
\toprule
\textbf{Stage} & \textbf{Agent Capability} & \textbf{Key Technologies} & \textbf{Human Role} & \textbf{Representative Systems} \\
\midrule
\textbf{I. Tool-Augmented} & Code completion, single-issue fixes, simple script generation & In-context learning, RAG & Author + reviewer & GitHub Copilot, Claude Code \\
\textbf{II. Single-Task Autonomous} & End-to-end feature building, debugging, basic system maintenance & Planning + tool use, self-correction & Intent architect + auditor & Devin, OpenHands \\
\textbf{III. Multi-Agent Teams} & Coordinated swarms for large systems, full-lifecycle management & Shared memory, role specialization, orchestration & PM + architect + auditor & LangChain orchestration, MetaGPT \cite{hong2024metagpt} \\
\textbf{IV. Self-Evolving Ecosystems} & Autonomous discovery, learning, reproduction, adaptation & Meta-learning, self-modification, ecosystem governance & Goal setter + ethics governor & AGI assistants (prospective) \\
\bottomrule
\end{tabular}
\end{table}

\subsection{Stage I: Tool-Augmented (2023--2025)}

The current dominant mode. Agents serve as assistants within human-led workflows. The breakthrough has been in coding: models can generate, explain, and debug code at near-expert level for well-scoped tasks. The limitation is that the human must still decompose problems, design architecture, and verify correctness.

\subsection{Stage II: Single-Task Autonomous (2025--2027)}

Agents begin to own complete tasks from specification to deployment. Systems like Devin and OpenHands demonstrate that agents can autonomously navigate codebases, implement features, and submit pull requests. The human shifts from ``doing'' to ``specifying what to do and verifying what was done.''

\subsection{Stage III: Multi-Agent Teams (2026--2029)}

Specialized agents coordinate as teams, mirroring human engineering organizations. A ``product manager agent'' translates business requirements into technical specifications; ``architect agents'' design system structure; ``developer agents'' implement components; ``QA agents'' test and validate. Shared memory and observability become critical infrastructure. The LangChain pilot \cite{kumar2026agentic} represents an early validation of this pattern.

\subsection{Stage IV: Self-Evolving Ecosystems (2028+)}

Agents gain the ability to improve their own architectures, spawn specialized sub-agents for new problem domains, and adapt to environmental changes without human intervention. At this stage, the distinction between ``software'' and ``agent'' dissolves entirely---the agent \textit{is} the system, and it evolves continuously. Human involvement shifts to meta-level governance: setting ethical boundaries, defining value functions, and ensuring alignment.

\section{Implications and Recommendations}

\subsection{For Practitioners}

The transition to agentic engineering demands a deliberate re-skilling strategy:

\begin{enumerate}[leftmargin=*]
    \item \textbf{Shift from code production to intent engineering.} The most valuable skill is no longer writing code efficiently but articulating tasks with sufficient clarity, context, and constraints that agents can execute them correctly.
    
    \item \textbf{Build agent orchestration competence.} Understanding how to decompose work across agents, manage shared memory, and design evaluation rubrics will differentiate effective practitioners.
    
    \item \textbf{Invest in observability infrastructure.} Agent systems require fundamentally different monitoring than traditional software. Tracing an agent's reasoning chain, detecting hallucinations, and measuring outcome quality demand new tooling.
    
    \item \textbf{Adopt a ``human-in-the-loop, agent-in-the-driver's-seat'' posture.} The most effective model today is neither fully autonomous nor fully human-driven. Agents should own execution; humans should own intent, critical judgment, and ethical oversight.
\end{enumerate}

\subsection{For Researchers}

Several open problems emerge with particular urgency:

\begin{enumerate}[leftmargin=*]
    \item \textbf{Long-context state management.} As EvoClaw demonstrates, agents lose coherence over extended development sequences. Architectures for compressing, indexing, and retrieving relevant context at scale are critical.
    
    \item \textbf{Verification in open-ended settings.} Current benchmarks test isolated correctness; real-world systems require guarantees of safety, reliability, and maintainability over time. New verification frameworks that capture these temporal dimensions are needed.
    
    \item \textbf{Agent alignment at scale.} As agents become more autonomous and are composed into teams, ensuring that their collective behavior aligns with human values becomes both more important and more difficult.
    
    \item \textbf{Economic models.} How should agentic services be priced? Outcome-based pricing (per resolved issue, per deployed feature) may replace subscription and usage-based models, but the incentive structures and risk allocation need careful analysis.
\end{enumerate}

\subsection{For Organizations}

Organizations should begin preparing for the agentic transition now:

\begin{enumerate}[leftmargin=*]
    \item \textbf{Identify agent-ready workflows.} Not all software work is equally amenable to agent automation. Tasks with clear success criteria, well-defined scope, and existing test infrastructure are ideal starting points.
    
    \item \textbf{Invest in evaluation frameworks.} The quality of agent output depends critically on the quality of the evaluation signal. Organizations should build test suites that go beyond correctness to measure robustness, maintainability, and alignment with business intent.
    
    \item \textbf{Redesign team structures.} As individual productivity multiplies through agent leverage, team topologies must evolve. Smaller teams of ``agent orchestrators'' may replace larger teams of developers, with corresponding shifts in hiring, promotion, and career development.
\end{enumerate}

\section{Conclusion}

This paper has argued that the emergence of AI agents constitutes a paradigm shift in what software is, not merely how it is built. The transition from ``AI $\rightarrow$ Software $\rightarrow$ Result'' to ``Agent $\rightarrow$ Result'' does not eliminate software -- the agent is software, but of a fundamentally different kind: one where decision logic is generated at runtime rather than pre-encoded, and where the system can evolve without human intervention. This is a natural progression of the software engineering discipline, not its termination.

The shift is grounded in first principles. Traditional software requires human engineers to encode all decision logic explicitly; the complexity of this task grows exponentially with system size while human capacity remains fixed. Agentic software outsources decision-making to LLMs whose capacity scales with training compute, decoupling solution capability from human cognitive limits. This is a qualitative change in what kinds of problems become economically tractable.

Yet we are still in the early stages. Benchmarks like EvoClaw reveal a stark gap between isolated-task performance and sustained autonomous development. The current moment calls for ambitious but calibrated investment: embrace agentic software as a new paradigm while recognizing that fully autonomous systems remain a multi-year research challenge.

Agentic Engineering represents a fundamental expansion of the software engineering discipline. Its practitioners are not programmers who learned new tools but a new kind of professional: intent architects who direct AI agents toward complex outcomes. The old software engineering is not ending; it is growing into something larger.

\section*{Acknowledgments}
The author thanks the open-source community for making research artifacts and benchmarks publicly available, and the teams behind SWE-bench, EvoClaw, and LangChain for their foundational contributions to agent evaluation infrastructure.

\bibliographystyle{plain}

\end{document}